\documentclass[prd,draft,amsfonts,amssymb,amsmath,showpacs]{revtex4}

\begin{document}

\title{Dynamical Chiral Symmetry Breaking, Color Superconductivity, and Bose-Einstein Condensation
in an $SU(N_{c})\times U(N_{f})_{L}\times U(N_{f})_{R}$-invariant Supersymmetric Nambu$-$Jona-Lasinio Model 
at finite Temperature and Density}
\author{Tadafumi Ohsaku}
\affiliation{Institut f\"{u}r Theoretische Physik, Universit\"{a}t zu K\"{o}ln, 50937 K\"{o}ln, Germany}

\date{\today}

%\maketitle

\newcommand{\bmx}{\mbox{\boldmath $x$}}
\newcommand{\bmy}{\mbox{\boldmath $y$}}
\newcommand{\bmk}{\mbox{\boldmath $k$}}
\newcommand{\bmp}{\mbox{\boldmath $p$}}
\newcommand{\bmq}{\mbox{\boldmath $q$}}
\newcommand{\bmP}{\mbox{\boldmath $P$}}  
\newcommand{\kfey}{\ooalign{\hfil/\hfil\crcr$k$}}
\newcommand{\pfey}{\ooalign{\hfil/\hfil\crcr$p$}}
\newcommand{\qfey}{\ooalign{\hfil/\hfil\crcr$q$}}
\newcommand{\Deltafey}{\ooalign{\hfil/\hfil\crcr$\Delta$}}
\newcommand{\nablafey}{\ooalign{\hfil/\hfil\crcr$\nabla$}}
\newcommand{\Dfey}{\ooalign{\hfil/\hfil\crcr$D$}}
\newcommand{\partfey}{\ooalign{\hfil/\hfil\crcr$\partial$}}
\def\sech{\mathop{\rm sech}\nolimits}

\begin{abstract}

We investigate the phenomena of the dynamical chiral symmetry breaking ( DCSB ), 
color superconductivity ( CSC ),
and Bose-Einstein condensation ( BEC ) in a supersymmetric ( SUSY ) 
vector-like $SU(N_{c})$ gauge model at finite temperature and density.
Both the ${\cal N}=1$ four-dimensional and ${\cal N}=2$ three-dimensional cases are considered.
We employ the ${\cal N}=1$ four-dimensional generalized SUSY Nambu$-$Jona-Lasinio model
( ${\cal N}=1$ generalized ${\rm SNJL}_{4}$ ) 
with a chemical potential as the model Lagrangian.
The ${\cal N}=2$ three-dimensional theory is obtained by a simple dimensional reduction scheme
of the four-dimensional counterpart. 
In order to realize the DCSB and BCS-type CSC in this model, we introduce a SUSY soft mass term. 
After adopting the method of SUSY auxiliary fields with 
the Fierz transformation in color and flavor spaces, 
we discuss several possible breaking schemes of the global symmetries of the model.
The integrations of the auxiliary fields of composites in the effective potential are performed by using the steepest descent approximation.
Under the finite-temperature Matsubara formalism, the gap equations are derived and solved.
The roles of both the boson and fermion sectors in the BEC, DCSB and CSC
are examined by the quasiparticle excitation spectra and the gap equations. 
The physical properties of the DCSB and the CSC are studied in detail.
Examination on the physical property of the BEC is beyond scope of this paper.
It is found that the BEC, DCSB and CSC can coexist under a condition of model parameters.
Several important observations are obtained in the method of the construction of the
SUSY BCS-type theory, starting from a SUSY field theoretical framework.

\end{abstract}

\pacs{11.30.Pb, 11.30.Qc, 11.30.Rd, 74.20.Fg}

\maketitle

\section{Introduction}

Based on the successes of the explanations for experiments of superconductivity,
the method of BCS theory ( Bardeen-Cooper-Schrieffer )~[1] has widely influenced in theoretical physics until now.
Through the work of Nambu and Jona-Lasinio ( NJL ), which established 
the concept of spontaneous symmetry breaking~[2],
the BCS-NJL mechanism was applied to particles~[3,4,5], nuclei and hadrons~[6,7], and condensed matter physics.
In this framework, particle and/or hadron masses are interpreted as results of dynamical chiral symmetry breaking ( DCSB ). 
Recently, theory of color superconductivity ( CSC ) in quantum chromodynamics ( QCD ) 
has been extensively studied~[6].
In high-density QCD, quarks at the Fermi surface have very high momenta,
and $SU(3_{c})$ color gauge coupling becomes weak, the gauge interactions between quarks could be 
satisfactorily handled by perturbation theory at $\epsilon_{F}>\Lambda_{QCD}$~[6] 
( $\epsilon_{F}$; the Fermi energy, $\Lambda_{QCD}$; the renormalization-group-invariant infrared energy scale of QCD ):
The dominant interaction between quarks is provided by a one-gluon exchange.
Several QCD-motivated NJL-type models are used as an effective theory of QCD to investigate hadron physics, 
meson and diquark mass spectra, and CSC until now~[6,7]. 
The theory of CSC motivates us to have interests on several cold dense matters,
with promoting developments of the methods of field theory at finite temperature and density. 
In such kind of problems, it might be useful for us to employ 
both the methods of group-theoretical classifications of possible states ( orderings ), 
and microscopic theories of dynamics of elementary fields of systems.
Microscopic theories are parametrized by several parameters, and they might reflect a hierarchical structure of Nature.

\vspace{3mm}

Let us assume the existence of supersymmetry ( SUSY ) in the universe.
Consequently, all of known elementary particles have their superpartners.
Phenomenologically, if the ${\cal N}=1$ SUSY exists, 
it is a clear fact that it is spontaneously broken in the universe.
The modern particle theory supposes that the superpartners might have their masses in the TeV energy region~[8],
and now it is a target of the LHC experiment.
A natural question arises that, what will happen when a condensed matter has a very high-density
and its chemical potential or the Fermi energy ( the characteristic energy scale of a system )
reach at the TeV energy scale.
In such a problem, both particle theory and condensed matter theory should be employed. 
The purpose of this paper is to examine physics, especially the DCSB and (color)superconductivity
in a SUSY condensed matter with the Fermi energy comparable with masses of superpartners.

\vspace{3mm}

Recently, the investigations of CSC from the viewpoint of the SUSY gauge theories were given 
in literature~[9,10,11].
Reference~[9] discussed the symmetry breaking patterns of the ${\cal N}=1$ SQCD ( supersymmetric QCD ) 
with a nonzero chemical potential and a SUSY breaking mass,
according as the results of the nonperturbative method of SUSY gauge theories~[12]. 
An extension of Ref.~[9] to ${\cal N}=2$ case was given in Ref.~[11].
The authors of these works intended to classify the possible breaking patterns of 
the gauge and flavor symmetries of QCD in CSC states, by utilizing the solutions of SQCD. 
However, any gauge-symmetry-breaking two-body pairs 
like diquarks were not treated in their models. 
The authors of Ref.~[10] proposed a toy model for giving diquark-like condensates.
They introduced an $SO(N)$ gauge interaction between quark superfields stronger than that of $SU(3_{c})$,
and then they argued that the $SO(N)$ gauge dynamics gives two-body pair condensations in their model.
This $SO(N)$ gauge interaction seems artificial. 
Compared with these works, our motivation and/or context of this paper are different.
In this paper, we will propose a method for constructing a BCS-NJL-type theory of $SU(N_{c})$ DCSB and CSC.
We intend to make our method parallel with the BCS-NJL theory.
We will study DCSB and CSC in a SUSY condensed matter system.
For this purpose, an SNJL ( supersymmetric Nambu$-$Jona-Lasinio )-like model Lagrangian 
suitable to describe a DCSB and a CSC is introduced. 
We consider the situation where we can neglect the $SU(N_{c})$ gauge dynamics,
and an attractive interaction is provided from an SNJL-type four-body interaction.
Historically, the SNJL model was first introduced to investigate the phenomenon of DCSB in SUSY field theory~[13,14],
and an $SU(2)\times U(1)$-gauged SNJL was applied to the top quark condensation theory 
of electroweak symmetry breaking in the minimal SUSY Standard Model~[4].
An $SU(N_{c})$ SNJL model was used to describe phenomena of phase transitions in the early universe~[15].

\vspace{3mm}

This paper is organized as follows.
In Sec. II, the formalism for our investigation is presented.
The SUSY auxiliary superfields of the composites are introduced. 
The effective potential and the gap equations for DCSB and CSC will be obtained by the method of steepest descent approximation.
We use the component field formalism, because it is convenient to introduce
the SUSY Nambu notation of superconductivity~[16,17,18,19,6], 
and also suitable for the method of the steepest descent for our evaluation of the one-loop effective potential.
The quasiparticle excitation energy spectra are examined in Sec. III.
Because we treat a model with a finite chemical potential, a Bose-Einstein condensation ( BEC ) 
can take place in zero modes of scalar fields~[9,16,20]. 
We will find the condition for the realization of BEC in our model,
and examine the relation between the BEC, DCSB and CSC.
Our examination on the BEC is only kinematically ( by looking at quasiparticle excitation energy spectra )
and a dynamical description of it is beyond scope of this paper.
Section IV is devoted for our study on the DCSB without CSC: We solve the gap equation for self-consistent chiral mass. 
Especially the critical coupling of the DCSB is studied. 
Both the ${\cal N}=1$ four-dimensional and ${\cal N}=2$ three-dimensional cases
are considered. In Sec. V, we examine a CSC of an $SU(3_{c})\times U(2_{f})_{L}\times U(2_{f})_{R}$ model
without DCSB through numerical calculations of the gap equation. 
The summary of this work, discussions, and issues for further investigations are presented in Sec. VI.

\section{Formalism}

In this section, we show the formalism for our investigation.
First, we introduce the following Lagrangian of an ${\cal N}=1$ 
$SU(N_{c})$-gauged generalized-SNJL model in four-dimensional spacetime:
\begin{eqnarray}
{\cal L} &=& \int d^{2}\theta \frac{1+m_{\tilde{g}}\theta^{2}}{16k_{c}(g^{c}_{0})^{2}} W^{\alpha}W_{\alpha} 
+ \int d^{2}\bar{\theta}\frac{1+m_{\tilde{g}}\bar{\theta}^{2}}{16k_{c}(g^{c}_{0})^{2}} \bar{W}_{\dot{\alpha}}\bar{W}^{\dot{\alpha}}
+\int d^{2}\theta d^{2}\bar{\theta}[(1-\Delta^{2}\theta^{2}\bar{\theta}^{2})K_{0}+K_{1}],  \nonumber \\
K_{0} &\equiv& Q^{\dagger}e^{V}Q+\tilde{Q}^{\dagger}e^{-V}\tilde{Q}, \nonumber \\
K_{1} &\equiv& G_{1}(Q^{\dagger}e^{V}Q)(\tilde{Q}^{\dagger}e^{-V}\tilde{Q})
+G_{2}(Q^{\dagger}e^{V}Q)(Q^{\dagger}e^{V}Q) 
+G_{3}(\tilde{Q}^{\dagger}e^{-V}\tilde{Q})(\tilde{Q}^{\dagger}e^{-V}\tilde{Q}). 
\end{eqnarray}
Here, $W_{\alpha}$ ( $\bar{W}_{\dot{\alpha}}$ ) is a chiral ( antichiral ) gauge field strength.
$Q$ ( $\tilde{Q}$ ) is an $N_{c}\times N_{f}$-matrix-valued chiral matter superfield 
belonging to the fundamental ( antifundamental ) representations of both $SU(N_{c})$ and $SU(N_{f})$.
The definitions are given in terms of component fields:
\begin{eqnarray}
Q = \phi(y)+\sqrt{2}\theta\psi(y)+\theta\theta F(y), \quad
\tilde{Q} = \tilde{\phi}(y)+\sqrt{2}\theta\tilde{\psi}(y)+\theta\theta\tilde{F}(y), \quad 
y^{\mu} = x^{\mu}+i\theta\sigma^{\mu}\bar{\theta}.
\end{eqnarray}
The chirality constraint 
$\overline{D}_{\dot{\alpha}}Q=\overline{D}_{\dot{\alpha}}\tilde{Q}=D_{\alpha}Q^{\dagger}=D_{\alpha}\tilde{Q}^{\dagger}=0$
( $D_{\alpha}\equiv \partial/\partial\theta^{\alpha}+i\sigma^{\mu}_{\alpha\dot{\alpha}}\bar{\theta}^{\dot{\alpha}}\partial^{x}_{\mu}$, 
$\overline{D}_{\dot{\alpha}}\equiv -\partial/\partial \bar{\theta}^{\dot{\alpha}}- i\theta^{\alpha}\sigma^{\mu}_{\alpha\dot{\alpha}}\partial^{x}_{\mu}$ ) 
is satisfied. $N_{c}\ge 3$ and $N_{f}\ge 2$ with $3N_{c}>N_{f}$ will be considered here.
In this paper, we follow the conventions for metric, gamma matrices, and spinor algebra 
given in the textbook of Wess and Bagger~[21].
$K_{0}+K_{1}$ is a K\"{a}hler potential, a real-valued function of the matter fields.
$K_{1}$ includes four-fermion interactions.
$V\equiv V^{A}\lambda_{A}$ ( $A=1, \cdots, N^{2}_{c}-1$ ) denotes a real vector multiplet. 
The gauge transformation is defined as follows:
\begin{eqnarray}
& & Q \to e^{-iG}Q, \quad Q^{\dagger} \to Q^{\dagger}e^{iG^{\dagger}}, \quad 
\tilde{Q} \to e^{iG}\tilde{Q}, \quad \tilde{Q}^{\dagger} \to \tilde{Q}^{\dagger}e^{-iG^{\dagger}}, \nonumber \\
& & e^{V} \to e^{-iG^{\dagger}}e^{V}e^{iG}, \quad e^{-V} \to e^{iG^{\dagger}}e^{-V}e^{-iG}, \quad
V \to V + i(G-G^{\dagger}), \quad G \equiv \sum^{N^{2}_{c}-1}_{A=1} G^{A}\lambda_{A}.
\end{eqnarray}
Similarly, the global flavor rotation will be given by
\begin{eqnarray}
Q \to e^{-i\Gamma}Q, \quad Q^{\dagger} \to Q^{\dagger}e^{i\Gamma^{\dagger}}, \quad 
\tilde{Q} \to e^{i\Gamma}\tilde{Q}, \quad \tilde{Q}^{\dagger} \to \tilde{Q}^{\dagger}e^{-i\Gamma^{\dagger}}, \quad 
\Gamma \equiv \sum^{N^{2}_{f}-1}_{I=1}\Gamma^{I}\Upsilon_{I}.
\end{eqnarray}
The Hermitian generators $\lambda_{A}$ of $SU(N_{c})$, and $\Upsilon_{I}$ of $SU(N_{f})$ 
( $I=1, \cdots, N^{2}_{f}-1$ ) are in the fundamental representations, and satisfy
\begin{eqnarray}
{\rm tr}\lambda_{A}\lambda_{B} = k_{c}\delta_{AB}, \quad {\rm tr}\Upsilon_{I}\Upsilon_{J} = k_{f}\delta_{IJ}, \quad
\sum^{N^{2}_{c}-1}_{A=1}(\lambda_{A}\lambda_{A})_{ij} = \frac{N^{2}_{c}-1}{2N_{c}}\delta_{ij}, \quad \sum^{N^{2}_{f}-1}_{I=1}(\Upsilon_{I}\Upsilon_{I})_{ij} = \frac{N^{2}_{f}-1}{2N_{f}}\delta_{ij}.
\end{eqnarray}
Here, $k_{c}$ and $k_{f}$ are normalization factors taken as $k_{c}=k_{f}=1/2$.
For the completeness relations, we should also include 
both $\lambda_{0}\equiv 1_{c}/\sqrt{2N_{c}}$ and $\Upsilon_{0}\equiv 1_{f}/\sqrt{2N_{f}}$
( the normalized unit matrices in color and flavor spaces ). 
Several Abelian symmetries are determined as follows:
\begin{eqnarray}
U(1)_{B}: \quad Q \to e^{i\alpha_{b}}Q, \quad \tilde{Q} \to e^{-i\alpha_{b}}\tilde{Q},  \qquad
U(1)_{A}: \quad Q \to e^{i\alpha_{a}}Q, \quad \tilde{Q} \to e^{i\alpha_{a}}\tilde{Q}.
\end{eqnarray}
$U(1)_{B}$ is a baryon number symmetry, $U(1)_{A}$ is an axial symmetry. 
The theory at the classical level has the following symmetries of unitary groups:
\begin{eqnarray}
SU(N_{c}) \times SU(N_{f})_{L} \times SU(N_{f})_{R} \times U(1)_{A} \times U(1)_{B},
\end{eqnarray}
( $=SU(N_{c}) \times [SU(N_{f})\times U(1)]_{L} \times [SU(N_{f})\times U(1)]_{R}$ ),
and the quantum numbers of the matter fields are summarized as follows:
\begin{eqnarray}
Q: \quad \Bigl( N_{c}, N_{f}, 1, 1, 1 \Bigr), \qquad 
\tilde{Q}: \quad \Bigl( \overline{N}_{c}, 1, \overline{N}_{f}, 1, -1 \Bigr).
\end{eqnarray}
A chemical potential $\mu$ is introduced by
\begin{eqnarray}
V &\to& V+2\mu\bar{\theta}\sigma^{0}\theta\otimes \lambda_{0}\otimes \Upsilon_{0}.
\end{eqnarray} 
Namely, $\mu$ acts as a zeroth-component of vector associated with the conserved charge of $U(1)_{B}$~[9].
With this form, all of the scalar $\phi,\tilde{\phi}$ and spinor $\psi,\tilde{\psi}$ fields share the same $\mu$.
$g^{c}_{0}$ is a bare gauge coupling constant defined at a cutoff energy scale.
$G_{1}$, $G_{2}$ and $G_{3}$ are coupling constants, and they have a mass dimension $[{\rm Mass}]^{-2}$.
Our model is non-renormalizable due to $K_{1}$.
In this model, we have introduced a holomorphic soft SUSY breaking mass $m_{\tilde{g}}$ for gauginos 
and a non-holomorphic universal soft mass $\Delta$ for squarks~[22].
Obviously, these soft masses do not break the symmetries given in (7).
An $R$-symmetry $\theta\to e^{i\alpha_{r}}\theta$ is explicitly broken by including the gaugino soft mass.

\vspace{3mm}

When $m_{\tilde{g}}=\Delta=G_{1}=G_{2}=G_{3}=0$, our model becomes SQCD,
while QCD will be obtained by $m_{\tilde{g}}\to\infty$, $\Delta\to\infty$, $G_{1}=G_{2}=G_{3}=0$.
By complete neglection of gauge fields, model (1) becomes a generalized-SNJL
( the original version of the SNJL model will be obtained by $G_{2}=G_{3}=0$ ), 
and NJL model will be derived by taking $\Delta\to\infty$ ( $\Delta\gg \Lambda$ ) in the SNJL.
Therefore, our $SU(N_{c})$-gauged SNJL model (1) might have a rich phase structure.
The vacua of SQCD were globally determined~[12]: SQCD has $D$-flat directions and the gauge group 
is broken partially or totally along these directions. 
The origin of the classical moduli space where all of expectation values of squarks vanish,
is a singular point and the gauge group is totally symmetric.  
In this paper, we consider the following situation: 
Starting from the origin of the classical moduli space of SQCD, we turn on the interaction of $K_{1}$.
Our model has different K\"{a}hler metric and Killing potential from that of SQCD.
However, the classical moduli space of our model is still the same with SQCD because 
they share the same condition for $D$-flat directions. 
Now, we take both the soft masses and chemical potential finite, and then we consider a quantum theory.
In the case when the one-loop beta function coefficient $3N_{c}-N_{f}$ 
is positive, the $SU(N_{c})$ gauge interaction becomes sufficiently weak 
from both the asymptotic freedom in a high-momentum region and a screening effect of
a finite-density medium, compared with the interactions arising from $K_{1}$.
We should also mention that the Cooper instability of the Fermi surface occurs with any attractive interaction.
The quantum dynamics of the system is dominated by the attractive interaction of $K_{1}$, 
and we restrict ourselves on the examination of it.
This situation might be achieved by a sufficiently large chemical potential $\mu$ with the following hierarchy,
\begin{eqnarray}
m_{\tilde{g}}\gg \Lambda > \mu\sim{\cal O}(\Delta) > \Lambda_{(S)QCD}, 
\end{eqnarray}
where, $\Lambda$ is a cutoff for the interactions in $K_{1}$.
$\Lambda_{(S)QCD}$ should be determined in SQCD with softly broken ${\cal N}=1$ SUSY~[22].
Under the condition (10), any intermediate processes of the gaugino field do not contribute in 
the dynamics of our theory.
Then the infrared phenomena of the SQCD itself, namely instanton effect, gaugino condensation, etc.,
will be irrelevant. Hence, we do not consider the nonperturbative generation of 
the Affleck-Dine-Seiberg superpotential at $N_{f}<N_{c}$ in our theory~[12]. 
We skip the question on the origin of the interaction of $K_{1}$.
For example, we could imagine that the attractive interaction of $K_{1}$ arised
from the one-gluon exchange interaction in SQCD with the condition (10)
( with the complete neglection of gaugino contributions ),
though a serious investigation for it demands us another work.
Because of the nonrenormalization theorem,
the chiral symmetry cannot be broken dynamically if the theory maintains the ${\cal N}=1$ SUSY 
exactly in the ordinary SNJL model~[13,14].
We speculate it is the case in our ${\cal N}=1$ model, and also in the ${\cal N}=2$ three-dimensional theory
obtained by a dimensional reduction. 
After we introduced $\Delta$, the realization of the DCSB and CSC through the BCS-NJL mechanism 
in this model might be possible through the dynamics governed by $K_{1}$.
Modern particle phenomenology considers $\Delta > $ TeV, and it is interesting for us to imagine the situation
discussed in the introduction of this paper.
The first term of $K_{1}$ might give a dynamically generated Dirac mass,
while a left-handed and a right-handed Majorana masses might be generated
by the second and third terms, respectively.
These terms include the factor $e^{\pm V}$ to show the $SU(N_{c})$ gauge invariance of the theory obviously.
The special case $G_{2}=G_{3}$ ( the left-right symmetric case ) will be taken 
to keep the parity symmetry in our model Lagrangian.
In fact, the spin-singlet ( Lorentz-scalar ) BCS pairing gap is given by a parity-invariant combination of 
a left-handed and a right-handed Majorana masses~[18,19].

\vspace{3mm}

After dropping the gauge fields, we apply the Fierz transformation to both the color and flavor spaces:
\begin{eqnarray}
(Q^{\dagger}Q)(\tilde{Q}\tilde{Q}^{\dagger}) &=& \frac{1}{k_{c}k_{f}}\sum^{N^{2}_{c}-1}_{A=0}\sum^{N^{2}_{f}-1}_{I=0}(Q^{\dagger}\lambda_{A}\Upsilon_{I}\tilde{Q}^{\dagger})(\tilde{Q}\lambda_{A}\Upsilon_{I}Q), \\
(Q^{\dagger}Q)(QQ^{\dagger}) &=& \frac{1}{k_{c}k_{f}}\sum^{N^{2}_{c}-1}_{B=0}\sum^{N^{2}_{f}-1}_{J=0}(Q^{\dagger}\lambda_{B}\Upsilon_{J}Q^{\dagger})(Q\lambda_{B}\Upsilon_{J}Q), \\
(\tilde{Q}^{\dagger}\tilde{Q})(\tilde{Q}\tilde{Q}^{\dagger}) &=& \frac{1}{k_{c}k_{f}}\sum^{N^{2}_{c}-1}_{C=0}\sum^{N^{2}_{f}-1}_{K=0}(\tilde{Q}^{\dagger}\lambda_{C}\Upsilon_{K}\tilde{Q}^{\dagger})(\tilde{Q}\lambda_{C}\Upsilon_{K}\tilde{Q}).
\end{eqnarray}
For example, the numbers of irreducible representations in the color space are estimated by
$N_{c}\otimes \overline{N_{c}} = 1\oplus N^{2}_{c}-1$, and
$N_{c}\otimes N_{c}=\frac{1}{2}N_{c}(N_{c}-1)\oplus \frac{1}{2}N_{c}(N_{c}+1)$.
Here, $\frac{1}{2}N_{c}(N_{c}-1)$ and $\frac{1}{2}N_{c}(N_{c}+1)$ give dimensions of antisymmetric and symmetric representations, respectively.
Consequently, our model Lagrangian will take the following form 
through the method of SUSY auxiliary fields~[4,14,15]:
\begin{eqnarray}
{\cal L} &=& \int d^{2}\theta d^{2}\bar{\theta}\Bigg[ (1-\Delta^{2}\theta^{2}\bar{\theta}^{2})
(Q^{\dagger}e^{2\mu\bar{\theta}\sigma^{0}\theta}Q+\tilde{Q}^{\dagger}e^{-2\mu\bar{\theta}\sigma^{0}\theta}\tilde{Q}) \nonumber \\
&& + \sum_{A,I}\Bigl\{ \frac{k_{c}k_{f}}{G_{1}}H^{\dagger}_{(1)AI}H_{(1)AI} + \delta(\bar{\theta})S_{(1)AI}\bigl( \frac{k_{c}k_{f}}{G_{1}}H_{(1)AI}
-\tilde{Q}\lambda_{A}\Upsilon_{I}Q \bigr)
 + \delta(\theta)S^{\dagger}_{(1)AI}\bigl( \frac{k_{c}k_{f}}{G_{1}}H^{\dagger}_{(1)AI}
-Q^{\dagger}\lambda_{A}\Upsilon_{I}\tilde{Q}^{\dagger} \bigr) \Bigr\}  \nonumber \\
&& + \sum_{B,J}\Bigl\{ \frac{k_{c}k_{f}}{G_{2}}H^{\dagger}_{(2)BJ}H_{(2)BJ} + \delta(\bar{\theta})S_{(2)BJ}\bigl( \frac{k_{c}k_{f}}{G_{2}}H_{(2)BJ}
-Q\lambda_{B}\Upsilon_{J}Q \bigr) 
 + \delta(\theta)S^{\dagger}_{(2)BJ}\bigl( \frac{k_{c}k_{f}}{G_{2}}H^{\dagger}_{(2)BJ}
-Q^{\dagger}\lambda_{B}\Upsilon_{J}Q^{\dagger} \bigr) \Bigr\}  \nonumber \\
&& + \sum_{C,K}\Bigl\{ \frac{k_{c}k_{f}}{G_{3}}H^{\dagger}_{(3)CK}H_{(3)CK} + \delta(\bar{\theta})S_{(3)CK}\bigl( \frac{k_{c}k_{f}}{G_{3}}H_{(3)CK}
-\tilde{Q}\lambda_{C}\Upsilon_{K}\tilde{Q} \bigr) 
 + \delta(\theta)S^{\dagger}_{(3)CK}\bigl( \frac{k_{c}k_{f}}{G_{3}}H^{\dagger}_{(3)CK}
-\tilde{Q}^{\dagger}\lambda_{C}\Upsilon_{K}\tilde{Q}^{\dagger} \bigr) \Bigr\}  \Bigg].  \nonumber \\
& & 
\end{eqnarray}
$S_{(1)AI}$, $S_{(2)BJ}$ and $S_{(3)CK}$ are SUSY Lagrange multipliers to keep the relations:
\begin{eqnarray}
H_{(1)AI} = \frac{G_{1}}{k_{c}k_{f}}\tilde{Q}\lambda_{A}\Upsilon_{I}Q, \quad 
H_{(2)BJ} = \frac{G_{2}}{k_{c}k_{f}}Q\lambda_{B}\Upsilon_{J}Q, \quad
H_{(3)CK} = \frac{G_{3}}{k_{c}k_{f}}\tilde{Q}\lambda_{C}\Upsilon_{K}\tilde{Q}. 
\end{eqnarray}
In terms of the lowest component of superfields, these relations give
$\phi_{H_{(1)AI}}=\frac{G_{1}}{k_{c}k_{f}}\tilde{\phi}\lambda_{A}\Upsilon_{I}\phi, \cdots$, etc.
Hence, $\phi_{H_{(1)AI}},\cdots,$ are given as squark pair condensates.
On the other hand, the scalars of $S_{(1)AI}, \cdots$ do not have such compact expressions 
they might show their "compositeness" in terms of components of $Q$ and $\tilde{Q}$.
Because the superfields $Q$ and $\tilde{Q}$ are bosonic, 
$Q\lambda_{B}\Upsilon_{J}Q$ and $\tilde{Q}\lambda_{C}\Upsilon_{K}\tilde{Q}$ 
have to be totally symmetric by transpositions.

\vspace{3mm}

To examine a dynamical breakdown of symmetries in the Lagrangian (14), 
we have to integrate out the matter fields $Q$ and $\tilde{Q}$ to obtain a "low-energy effective action", the SUSY Ginzburg-Landau ( GL ) 
functional given in terms of the SUSY auxiliary fields, similar to the well-known procedure 
of the theory of superconductivity. 
The Nambu-Goldstone scalars and spinors can be examined by the SUSY GL functional.
At zero-temperature and far below the compositeness energy scale, 
the SUSY composite auxiliary fields in (14) appear as independent dynamical degrees of freedom,
with having radiatively generated kinetic terms ( precisely, K\"{a}hler potentials ) of 
the Lagrange multipliers $S_{(1)AI}, \cdots$ ~[4].
While, a finite-temperature effect excites quasiparticles over a BCS energy gap, 
and they might feel dissociation of condensation pairs.
The method of the nonperturbative holomorphic effective action~[12] 
could be applied to a SUSY GL functional.
( In this context, the method of the nonperturbative holomorphic effective action 
seems to be a kind of group-theoretical classification of the GL theory in superconductivity. )
Several possible solutions of it might be classified by this method,
however, this way will give a complicated situation as we have mentioned in the introduction~[10].
In this paper, we use the method of steepest descent, a kind of the Hartree-Fock mean field theory,
for the integrations of the auxiliary fields of composites.
In fact, at least in condensed matter physics of four-dimensional spacetime, 
the mean field approximation such as the BCS treatment provides a nice description for superconductivity.
The success of the BCS method is one of the bases of the investigations of the theory of CSC.  
We consider it is also the case in our theory in four dimensional spacetime.
In three-dimensional case, a mean field description for a phase transition becomes bad because of the existence of 
strong quantum fluctuations, and usually a Kosterlitz-Thouless type transition occurs~[23].    
However, a treatment of the BCS mean field level will be a starting point 
of further investigations on superconductivity in three dimensions.
Expanding the Lagrangian (14) in terms of the component fields, 
eliminating the auxiliary fields of chiral multiplets through their Euler equations, 
and keeping only the relevant terms in the steepest descent approximation, we get
\begin{eqnarray}
{\cal L} &=& -k_{c}k_{f}\Bigl( \sum_{A,I}\frac{|\phi_{S_{(1)AI}}|^{2}}{G_{1}}+ \sum_{B,J}\frac{|\phi_{S_{(2)BJ}}|^{2}}{G_{2}}+ \sum_{C,K}\frac{|\phi_{S_{(3)CK}}|^{2}}{G_{3}} \Bigr) \nonumber \\
& & -(\partial_{\nu}-i\mu\delta_{\nu 0})\phi^{\dagger}(\partial^{\nu}+i\mu\delta_{\nu 0})\phi
    -(\partial_{\nu}+i\mu\delta_{\nu 0})\tilde{\phi}^{\dagger}(\partial^{\nu}-i\mu\delta_{\nu 0})\tilde{\phi} \nonumber \\
& & -\phi^{\dagger}(\Delta^{2}+M^{B}_{1})\phi
    -\tilde{\phi}^{\dagger}(\Delta^{2}+M^{B}_{1})\tilde{\phi}  
    +\phi^{\dagger}M^{B}_{2}\tilde{\phi} +\tilde{\phi}^{\dagger}M^{B\dagger}_{2}\phi \nonumber \\
& & -i\bar{\psi}\bar{\sigma}^{\nu}(\partial_{\nu}-i\mu\delta_{\nu 0})\psi
    -i\bar{\tilde{\psi}}\bar{\sigma}^{\nu}(\partial_{\nu}+i\mu\delta_{\nu 0})\tilde{\psi}   \nonumber \\
& & + \tilde{\psi}M^{F}_{1}\psi 
    + \bar{\psi}M^{F\dagger}_{1}\bar{\tilde{\psi}} 
    + \psi M^{F}_{2}\psi 
    + \bar{\psi}M^{F\dagger}_{2}\bar{\psi} 
    + \tilde{\psi}M^{F}_{3}\tilde{\psi} 
    + \bar{\tilde{\psi}}M^{F\dagger}_{3}\bar{\tilde{\psi}}.
\end{eqnarray}
$\phi_{S_{(1)AI}}$, $\phi_{S_{(2)BJ}}$ and $\phi_{S_{(3)CK}}$  
denote the scalar components of the Lagrange multiplier multiplets. 
We also assumed that, all of scalars of the SUSY auxiliary fields are independent on spacetime coordinates
while all of the spinor components of them are zero.
The mass matrices are defined as follows: For scalar fields,
\begin{eqnarray}
M^{B}_{1} &\equiv& |\sum_{A,I}\phi_{S_{(1)AI}}\lambda_{A}\Upsilon_{I}|^{2}+4| \sum_{B,J}\phi_{S_{(2)BJ}}\lambda_{B}\Upsilon_{J}|^{2}, \nonumber \\
M^{B}_{2} &\equiv& 
-2\Bigl(\sum_{A,I}\phi_{S_{(1)AI}}\lambda_{A}\Upsilon_{I}\Bigr)\Bigl(\sum_{B,J}\phi^{\dagger}_{S_{(2)BJ}}\lambda_{B}\Upsilon_{J}\Bigr)
-2\Bigl(\sum_{C,K}\phi_{S_{(3)CK}}\lambda_{C}\Upsilon_{K}\Bigr)\Bigl(\sum_{A,I}\phi^{\dagger}_{S_{(1)AI}}\lambda_{A}\Upsilon_{I}\Bigr), \nonumber \\
& & 
\end{eqnarray}
and for spinor fields, 
\begin{eqnarray}
M^{F}_{1} \equiv  \sum_{A,I}\phi_{S_{(1)AI}}\lambda_{A}\Upsilon_{I}, \quad
M^{F}_{2} \equiv  \sum_{B,J}\phi_{S_{(2)BJ}}\lambda_{B}\Upsilon_{J}, \quad
M^{F}_{3} \equiv  \sum_{C,K}\phi_{S_{(3)CK}}\lambda_{C}\Upsilon_{K}.
\end{eqnarray}
The matrices $\psi M^{F}_{2}\psi$ and $\tilde{\psi}M^{F}_{3}\tilde{\psi}$ have to be totally antisymmetric
under the exchange of spinor, color and flavor indices because of the Pauli principle. 
This gives us some restrictions on the choices of the symmetries of the SUSY composite auxiliary fields.
Now, one finds the partition function as follows:
\begin{eqnarray}
{\cal Z} &=& \int 
{\cal D}\phi{\cal D}\phi^{\dagger}{\cal D}\tilde{\phi}{\cal D}\tilde{\phi}^{\dagger}{\cal D}\Psi{\cal D}\bar{\Psi}\prod_{A,I}{\cal D}\phi_{S_{(1)AI}}{\cal D}\phi^{\dagger}_{S_{(1)AI}}
\prod_{B,J}{\cal D}\phi_{S_{(2)BJ}}{\cal D}\phi^{\dagger}_{S_{(2)BJ}}
\prod_{C,K}{\cal D}\phi_{S_{(3)CK}}{\cal D}\phi^{\dagger}_{S_{(3)CK}} \nonumber \\
&& \times \exp\Bigg[ i\int d^{4}x \Bigl\{
-k_{c}k_{f}\bigl( \sum_{A,I}\frac{|\phi_{S_{(1)AI}}|^{2}}{G_{1}}+ \sum_{B,J}\frac{|\phi_{S_{(2)BJ}}|^{2}}{G_{2}}+ \sum_{C,K}\frac{|\phi_{S_{(3)CK}}|^{2}}{G_{3}} \bigr) +\Pi^{\dagger}\Omega_{B}(p_{\nu})\Pi + \frac{1}{2}\bar{\Xi}\Omega_{F}(p_{\nu})\Xi \Bigr\} \Bigg]  \nonumber \\
&=& \int \prod_{A,I}{\cal D}\phi_{S_{(1)AI}}{\cal D}\phi^{\dagger}_{S_{(1)AI}}
\prod_{B,J}{\cal D}\phi_{S_{(2)BJ}}{\cal D}\phi^{\dagger}_{S_{(2)BJ}}
\prod_{C,K}{\cal D}\phi_{S_{(3)CK}}{\cal D}\phi^{\dagger}_{S_{(3)CK}} \nonumber \\
&& \times \exp\Bigg[ i\int d^{4}x \Bigl\{ -k_{c}k_{f}(\sum_{A,I}\frac{|\phi_{S_{(1)AI}}|^{2}}{G_{1}}+\sum_{B,J}\frac{|\phi_{S_{(2)BJ}}|^{2}}{G_{2}}+\sum_{C,K}\frac{|\phi_{S_{(3)CK}}|^{2}}{G_{3}} ) \Bigr\} \nonumber \\
& & \qquad -V_{BEC} + 2i\ln{\rm Det}\Omega_{B} -i\ln{\rm Det}\Omega_{F}  \Bigg], \\
V_{BEC} &\equiv& -(\Pi^{c})^{\dagger}\Omega_{B}(p_{\nu}=0)\Pi^{c} = \phi^{\dagger}(\mu^{2}+M^{B}_{1}+\Delta^{2})\phi+\tilde{\phi}^{\dagger}(\mu^{2}+M^{B}_{1}+\Delta^{2})\tilde{\phi}-\phi^{\dagger}M^{B}_{2}\tilde{\phi}-\tilde{\phi}^{\dagger}M^{B\dagger}_{2}\phi.
\end{eqnarray}
Here, we have used the following definitions of the fields:
\begin{eqnarray}
\Pi \equiv 
\left( 
\begin{array}{c}
\phi \\
\tilde{\phi} 
\end{array} 
\right)=\Pi^{c}+\Pi^{p_{\nu}\ne 0}(x), 
\quad 
\Psi \equiv 
\left( 
\begin{array}{c}
\psi \\
\bar{\tilde{\psi}} 
\end{array} 
\right),
\quad
\Xi \equiv 
\left(
\begin{array}{c}
\Psi \\
\bar{\Psi}^{T}
\end{array}
\right), 
\quad 
\bar{\Xi} \equiv ( \bar{\Psi}, \Psi^{T} ).   
\end{eqnarray}
In the expression given above, the condensate $\Pi^{c}$ is separated from $\Pi$ 
for taking into account the possible BEC in scalar fields.
$\Pi^{c}$ does not depend on spacetime coordinates. $\Pi^{c}\ne 0$ when the BEC takes place, otherwise $\Pi^{c}=0$.
$V_{BEC}$ has no quartic term of $\Pi^{c}$, and describes the BEC instability of a vacuum. 
$\Psi$ takes the form of the four-component Dirac bispinor in the spinor space.
$\Xi$ and $\bar{\Xi}$ are the Nambu notations~[16,17,18], they have $8\times N_{c}\times N_{f}$ components.
$\Psi$ itself belongs to the fundamental representations of both $SU(N_{c})$ and $SU(N_{f})_{L+R}$.
The definitions of the matrices $\Omega_{B}$ and $\Omega_{F}$ are given as follows: 
\begin{eqnarray}
\Omega_{B}(p_{\nu}) &\equiv& \left(
\begin{array}{cccc}
(\partial_{\nu}+i\mu\delta_{\nu 0})(\partial^{\nu}+i\mu\delta_{\nu 0})-M^{B}_{1}-\Delta^{2} & M^{B}_{2}  \\
M^{B\dagger}_{2}  &  (\partial_{\nu}-i\mu\delta_{\nu 0})(\partial^{\nu}-i\mu\delta_{\nu 0})-M^{B}_{1}-\Delta^{2}
\end{array}
\right),  \\
\Omega_{F}(p_{\nu}) &\equiv& \left(
\begin{array}{cccc}
i\partfey+\gamma^{0}\mu-M^{F}_{1}\frac{1+i\gamma_{5}}{2}-M^{F\dagger}_{1}\frac{1-i\gamma_{5}}{2} & -M^{F}_{3}C(1+i\gamma_{5})-M^{F\dagger}_{2}C(1-i\gamma_{5}) \\
-M^{F}_{2}C(1+i\gamma_{5})-M^{F\dagger}_{3}C(1-i\gamma_{5}) & i\partfey^{T}-\gamma^{0T}\mu+M^{F}_{1}\frac{1+i\gamma_{5}}{2}+M^{F\dagger}_{1}\frac{1-i\gamma_{5}}{2}
\end{array}
\right), 
\end{eqnarray}
where, $\gamma_{5}\equiv \gamma^{0}\gamma^{1}\gamma^{2}\gamma^{3}$, 
and $C\equiv i\gamma^{2}\gamma^{0}$ is the charge conjugation matrix.
If we take only $\lambda_{0}$ and $\Upsilon_{0}$ in all of the expansions of the Fierz transformations (11)-(13),
we obtain the equivalent result with the $U(1)$-gauge case given in Ref.~[16].
The numbers of eigenvalues of $\Omega_{B}$ and $\Omega_{F}$ become
$4\times N_{c}\times N_{f}$ and $8\times N_{c}\times N_{f}$, respectively.

\vspace{3mm}

Next, we have to consider several possible breaking schemes for the global
$SU(N_{c}) \times SU(N_{f})_{L} \times SU(N_{f})_{R} \times U(1)_{A} \times U(1)_{B}$-symmetries of our theory. 
In the relativistic theory of superconductivity,
usually we take the parity-invariant Lorentz-scalar symmetry for the spin singlet BCS pairing state~[6,16,18,19].
In the case of QCD, scalar will be preferred than pseudoscalar due to the $U(1)_{A}$-breaking instanton effect.
Our model is not the same with QCD, though
we will choose the gap as an antisymmetric function in spinor space with keeping the parity invariance.
In order to take the symmetries of both the Dirac mass and the BCS pairing gap in $\Omega_{F}$ as scalar,
we choose 
\begin{eqnarray}
M^{F\dagger}_{1}=M^{F}_{1}, \qquad M^{F\dagger}_{2}=-M^{F}_{3}.
\end{eqnarray}
The gap function of $M^{F\dagger}_{2}\ne -M^{F}_{3}$ corresponds to a
linear combination of the scalar and pseudoscalar Cooper pairings.
We will drop the pseudoscalar by the constraint $M^{F\dagger}_{2}=-M^{F}_{3}$. 
In the theory of $SU(3_{c})$ CSC, the symmetry of pairing gap of quarks will be determined
from the consideration on the attractive channel of one-gluon exchange and the Pauli principle~[6].
Because we choose superconducting order parameters as antisymmetric ( i.e., a Lorentz scalar ) in spinor indices, 
there are two possible cases:
(a) symmetric in both color and flavor indices, (b) antisymmetric in both color and flavor indices.
The attractive channel of $SU(3_{c})$ is given in the antisymmetric $\bar{\bf 3}$ representation 
of the decomposition ${\bf 3}\otimes{\bf 3}=\bar{\bf 3}\oplus{\bf 6}$.
For example, in the case of $SU(3_{c})\times SU(2_{f})_{L}\times SU(2_{f})_{R}$, 
the symmetry of a Cooper pair is expressed as~[6]
\begin{eqnarray}
\langle q^{T}_{ai}(C\gamma_{5})q_{bj} \rangle \sim \epsilon_{ij}\epsilon_{abc}.
\end{eqnarray}
Here, $q$ is a Dirac bispinor of quark field. The quark pair is completely antisymmetric with respect 
to spinor, color ( $a,b,c$ ) and flavor ( $i,j$ ) indices. 
The index $c$ is taken arbitrarily in color space.
The color symmetry is broken as $SU(3_{c})\to SU(2_{c})\times U(1)_{\lambda_{8}}$.
At $N_{f}=3$, an order parameter of so-called color-flavor locked phase, 
where color and flavor degrees of freedom are coupled together, is energetically favorable:
\begin{eqnarray}
\langle q^{T}_{ai}(C\gamma_{5})q_{bj} \rangle \sim \sum^{N_{c}}_{I=1}\sum^{N_{f}}_{J=1}c_{IJ}\epsilon_{ijI}\epsilon_{abJ}.
\end{eqnarray}
Here, $c_{IJ}$ is determined variationally.
Such a color-flavor locked phase are generally favorable at $N_{f}\ge 3$ in $SU(3_{c})$ QCD~[24].

\vspace{3mm}

To derive the gap equations, we consider the breaking scheme of the special case $SU(3_{c})\times SU(2_{f})_{L}\times SU(2_{f})_{R}$,
the most established breaking scheme in theory of $SU(3_{c})$ CSC.
Because the chiral group $SU(2_{f})_{L}\times SU(2_{f})_{R}$ is unbroken by the CSC itself,
we can treat the DCSB and the CSC separately, 
and it greatly reduces the effort for our numerical calculations of gap equations.
We take only 
\begin{eqnarray}
\lambda_{A}=\lambda_{0}, \quad \lambda_{B}=\lambda_{C}=\lambda_{3}, \quad 
\Upsilon_{I}=\Upsilon_{0}, \quad \Upsilon_{B}=\Upsilon_{C}=\frac{\sigma_{2}}{2}
\end{eqnarray}
( the index $c$ in Eq. (25) is taken to the third direction in color space )
in Eq.(19), and include one $SU(3_{c})$-singlet meson.
Now, the Nambu notation $\Xi$ has $8\times 3\times 2=48$ components.
Hereafter, we use $\phi_{1}\equiv \phi_{S_{(1)00}}$ and $\phi_{2}\equiv \phi_{S_{(2)30}}$.
The boson and fermion determinants in Eq. (19) will be evaluated in the following forms:
\begin{eqnarray}
{\rm det}\Omega_{B} 
&=& \bigl[(p_{0}-E^{B}_{+}(\bmp))(p_{0}+E^{B}_{+}(\bmp))(p_{0}-E^{B}_{-}(\bmp))(p_{0}+E^{B}_{-}(\bmp))\bigr]^{2N_{f}} \nonumber \\
& & \times \bigl[(p_{0}-{\cal E}^{B}_{+}(\bmp))(p_{0}+{\cal E}^{B}_{+}(\bmp))(p_{0}-{\cal E}^{B}_{-}(\bmp))(p_{0}+{\cal E}^{B}_{-}(\bmp))]^{(N_{c}-2)N_{f}},  \\
{\rm det}\Omega_{F} 
&=& \bigl[(p_{0}-E^{F}_{+}(\bmp))(p_{0}+E^{F}_{+}(\bmp))(p_{0}-E^{F}_{-}(\bmp))(p_{0}+E^{F}_{-}(\bmp))\bigr]^{4N_{f}} \nonumber \\
& & \times \bigl[(p_{0}-{\cal E}^{F}_{+}(\bmp))(p_{0}+{\cal E}^{F}_{+}(\bmp))(p_{0}-{\cal E}^{F}_{-}(\bmp))(p_{0}+{\cal E}^{F}_{-}(\bmp))\bigr]^{2(N_{c}-2)N_{f}}.
\end{eqnarray}
All of the eigenvalues of $\Omega_{F}$ have two-fold degeneracies in spinor space, and
these degeneracies relate to the time-reversal invariance of the BCS gap function~[18].
The bosonic and fermionic quasiparticle energy spectra are evaluated to be,
\begin{eqnarray}
E^{B}_{+}(\bmp) &\equiv& \sqrt{ \bmp^{2}+|\phi_{1}|^{2}+4|\phi_{2}|^{2}+\Delta^{2}} - \mu,  \\
E^{B}_{-}(\bmp) &\equiv& \sqrt{ \bmp^{2}+|\phi_{1}|^{2}+4|\phi_{2}|^{2}+\Delta^{2}} + \mu,  \\
{\cal E}^{B}_{+}(\bmp) &\equiv& \sqrt{\bmp^{2}+|\phi_{1}|^{2}+\Delta^{2}}-\mu, \\
{\cal E}^{B}_{-}(\bmp) &\equiv& \sqrt{\bmp^{2}+|\phi_{1}|^{2}+\Delta^{2}}+\mu, \\
E^{F}_{+}(\bmp) &\equiv& \sqrt{(\sqrt{\bmp^{2}+|\phi_{1}|^{2}}-\mu)^{2}+4|\phi_{2}|^{2}},  \\
E^{F}_{-}(\bmp) &\equiv& \sqrt{(\sqrt{\bmp^{2}+|\phi_{1}|^{2}}+\mu)^{2}+4|\phi_{2}|^{2}},  \\
{\cal E}^{F}_{+}(\bmp) &\equiv& \sqrt{\bmp^{2}+|\phi_{1}|^{2}}-\mu,  \\
{\cal E}^{F}_{-}(\bmp) &\equiv& \sqrt{\bmp^{2}+|\phi_{1}|^{2}}+\mu.
\end{eqnarray}
From the forms of these spectra, 
we confirm $|\phi_{1}|$ is a dynamically generated Dirac mass, while $2|\phi_{2}|$ 
corresponds to a BCS gap function.
The quasiparticles of the branches ${\cal E}^{B}_{\pm}$, ${\cal E}^{F}_{\pm}$ do not participate
the BCS-type superconductivity.

\vspace{3mm}

As we have mentioned above, we will consider the case $G_{2}=G_{3}$.
The effective action under the steepest descent approximation becomes
\begin{eqnarray}
S_{eff} &=& \int d^{4}x\Bigl[ -V_{BEC} -\frac{|\phi_{1}|^{2}}{4G_{1}} -\frac{|\phi_{2}|^{2}}{2G_{2}}  \Bigr]+2i\ln{\rm Det}\Omega_{B} -i\ln{\rm Det}\Omega_{F},  \\
\frac{1}{2}V_{BEC} &=& (E^{B}_{+}(\bmp=0))^{2}\sum^{2N_{f}}_{i=1}|\phi^{c}_{i}|^{2}+(E^{B}_{-}(\bmp=0))^{2}\sum^{2N_{f}}_{j=1}|\phi^{c}_{j}|^{2}  \nonumber \\
& & +({\cal E}^{B}_{+}(\bmp=0))^{2}\sum^{(N_{c}-2)N_{f}}_{k=1}|\phi^{c}_{k}|^{2}+({\cal E}^{B}_{-}(\bmp=0))^{2}\sum^{(N_{c}-2)N_{f}}_{l=1}|\phi^{c}_{l}|^{2}, \\
& & \quad \Pi^{c} = 
\left( 
\begin{array}{cccc}
\phi^{c}_{i}, & \phi^{c}_{j}, & \phi^{c}_{k}, & \phi^{c}_{j} 
\end{array} 
\right)^{T}.
\end{eqnarray}
Here, $\Pi^{c}$ has been divided according to the breaking scheme in the color space.
From the form of $V_{BEC}$, we recognize the following fact: 
When one of the excitation energy spectra of scalar fields
has a zero at $\bmp=0$, the BEC takes place and at least one of components of $\Pi^{c}$ obtains 
a finite vacuum expectation value.
Now, we introduce the finite-temperature Matsubara formalism~[20] obtained by the following substitutions in our theory:
\begin{eqnarray}
\int\frac{dp_{0}}{2\pi i} \to \sum_{n}\frac{1}{\beta}, \qquad p_{0} \to i\omega^{B}_{n}, i\omega^{F}_{n},
\end{eqnarray}
where $\beta\equiv 1/k_{B}T$ ( $k_{B}$; the Boltzmann constant, $T$; temperature ). 
We take $k_{B}=1$ throughout this paper.
$\omega^{B}_{n}$ and $\omega^{F}_{n}$ are boson and fermion discrete frequencies, respectively.
Their definitions are $\omega^{B}_{n}\equiv 2n\pi/\beta$ and $\omega^{F}_{n}\equiv (2n+1)\pi/\beta$ 
( $n=0,\pm 1,\pm 2, \cdots$ ). 
The effective potential ( thermodynamic potential ) becomes
\begin{eqnarray}
&& V_{eff}(\phi_{1},\phi_{2}) = V_{BEC}+\frac{|\phi_{1}|^{2}}{4G_{1}} + \frac{|\phi_{2}|^{2}}{2G_{2}} + \sum_{n}\frac{1}{\beta}\int^{\Lambda}\frac{d^{3}\bmp}{(2\pi)^{3}}\ln\det\Omega_{F} -2\sum_{n}\frac{1}{\beta}\int^{\Lambda}\frac{d^{3}\bmp}{(2\pi)^{3}}\ln\det\Omega_{B}   \nonumber  \\
&=& V_{BEC}+\frac{|\phi_{1}|^{2}}{4G_{1}} + \frac{|\phi_{2}|^{2}}{2G_{2}} + \int^{\Lambda}\frac{d^{3}\bmp}{(2\pi)^{3}}  \nonumber \\
& & \times \Bigg( 2N_{f}(E^{B}_{+}(\bmp)+E^{B}_{-}(\bmp)-E^{F}_{+}(\bmp)-E^{F}_{-}(\bmp))
    +(N_{c}-2)N_{f}({\cal E}^{B}_{+}(\bmp)+{\cal E}^{B}_{-}(\bmp)-{\cal E}^{F}_{+}(\bmp)-{\cal E}^{F}_{-}(\bmp))    \nonumber \\ 
& & -\frac{4N_{f}}{\beta}\ln(1+e^{-\beta E^{F}_{+}(\bmp)})(1+e^{-\beta E^{F}_{-}(\bmp)})
    +\frac{4N_{f}}{\beta}\ln(1-e^{-\beta E^{B}_{+}(\bmp)})(1-e^{-\beta E^{B}_{-}(\bmp)}) \nonumber \\
& & -\frac{2(N_{c}-2)N_{f}}{\beta}\ln(1+e^{-\beta {\cal E}^{F}_{+}(\bmp)})(1+e^{-\beta {\cal E}^{F}_{-}(\bmp)})
    +\frac{2(N_{c}-2)N_{f}}{\beta}\ln(1-e^{-\beta {\cal E}^{B}_{+}(\bmp)})(1-e^{-\beta {\cal E}^{B}_{-}(\bmp)}) \Bigg). 
\end{eqnarray}
To obtain the final expression in Eq. (42), the frequency summations were performed. 
A three-dimensional momentum cutoff $\Lambda$ was introduced to regularize the integral.
The gap equations are derived in the following forms: 
\begin{eqnarray}
0 = \frac{\partial V_{eff}}{\partial |\phi_{1}| } 
&=& \frac{\partial V_{BEC}}{\partial |\phi_{1}| } + \frac{|\phi_{1}|}{2G_{1}} - 2N_{f}|\phi_{1}|\int^{\Lambda}\frac{d^{3}\bmp}{(2\pi)^{3}} \nonumber \\
& & \times \Bigg[ \bigl( 1-\frac{\mu}{\sqrt{\bmp^{2}+|\phi_{1}|^{2} } }\bigr)\frac{1}{E^{F}_{+}}\tanh\frac{\beta}{2}E^{F}_{+}+\bigl( 1+\frac{\mu}{\sqrt{\bmp^{2}+|\phi_{1}|^{2} } }\bigr)\frac{1}{E^{F}_{-}}\tanh\frac{\beta}{2}E^{F}_{-}   \nonumber \\
& & \quad - \frac{1}{\sqrt{\bmp^{2}+|\phi_{1}|^{2}+4|\phi_{2}|^{2}+\Delta^{2}}} 
\Bigl( \coth\frac{\beta}{2}E^{B}_{+} + \coth\frac{\beta}{2}E^{B}_{-} \Bigr)  \Bigg],  \nonumber \\
& & - (N_{c}-2)N_{f}|\phi_{1}|\int^{\Lambda}\frac{d^{3}\bmp}{(2\pi)^{3}}
\Bigg[ \frac{1}{\sqrt{\bmp^{2}+|\phi_{1}|^{2}}}\Bigl( \tanh\frac{\beta}{2}{\cal E}^{F}_{+} + \tanh\frac{\beta}{2}{\cal E}^{F}_{-} \Bigr) \nonumber \\
& & \qquad - \frac{1}{\sqrt{\bmp^{2}+|\phi_{1}|^{2}+\Delta^{2}}}\Bigl(\coth\frac{\beta}{2}{\cal E}^{B}_{+}+\coth\frac{\beta}{2}{\cal E}^{B}_{-} \Bigr)  \Bigg],
\end{eqnarray}
and,
\begin{eqnarray}
0 = \frac{\partial V_{eff}}{\partial |\phi_{2}| } 
&=& \frac{\partial V_{BEC}}{\partial |\phi_{2}| } +\frac{|\phi_{2}|}{G_{2}} - 8N_{f}|\phi_{2}|\int^{\Lambda}\frac{d^{3}\bmp}{(2\pi)^{3}}\Bigg[ \frac{1}{E^{F}_{+}}\tanh\frac{\beta}{2}E^{F}_{+}+\frac{1}{E^{F}_{-}}\tanh\frac{\beta}{2}E^{F}_{-}  \nonumber \\
& & \quad - \frac{1}{\sqrt{\bmp^{2}+|\phi_{1}|^{2}+4|\phi_{2}|^{2}+\Delta^{2}}}
\Bigl( \coth\frac{\beta}{2}E^{B}_{+} + \coth\frac{\beta}{2}E^{B}_{-} \Bigr) \Bigg].   
\end{eqnarray}
At first glance, both of these gap equations might have quadratic divergences, 
and the nature of the divergences might be influenced by $\Delta$. 
These gap equations correctly give their limiting cases at $\Delta\to \infty$.
For example, Eq. (44) gives the gap equation of the non-SUSY CSC~[6,18,19] at $\Delta\to \infty$.
Equations (43) and (44) can have nontrivial solutions at least at $G_{1},G_{2}>0$ ( attractive interactions ).
The baryon density $\varrho_{B}$, the conjugate of $\mu$, is found to be
\begin{eqnarray}
\varrho_{B} = -\frac{\partial V_{eff}}{\partial \mu} 
&=& -\frac{\partial V_{BEC}}{\partial\mu} +2N_{f} \int\frac{d^{3}\bmp}{(2\pi)^{3}} \Bigg( 
\frac{\mu-\sqrt{\bmp^{2}+|\phi_{1}|^{2}}}{E^{F}_{+}}\tanh\frac{\beta}{2}E^{F}_{+}+\frac{\mu+\sqrt{\bmp^{2}+|\phi_{1}|^{2}}}{E^{F}_{-}}\tanh\frac{\beta}{2}E^{F}_{-}  \nonumber \\
& & +\coth\frac{\beta}{2}E^{B}_{+} - \coth\frac{\beta}{2}E^{B}_{-}   \Bigg) \nonumber \\
& & +(N_{c}-2)N_{f}\int\frac{d^{3}\bmp}{(2\pi)^{3}} \Bigg( -\tanh\frac{\beta}{2}{\cal E}^{F}_{+}+\tanh\frac{\beta}{2}{\cal E}^{F}_{-} + \coth\frac{\beta}{2}{\cal E}^{B}_{+}-\coth\frac{\beta}{2}{\cal E}^{B}_{-} \Bigg).   
\end{eqnarray}
In the zero-temperature case with $|\phi_{2}|=0$ and $\Pi^{c}=0$, one has
\begin{eqnarray}
\varrho_{B} &=& 2N_{c}N_{f}\int \frac{d^{3}\bmp}{(2\pi)^{3}} \theta(\mu-\sqrt{\bmp^{2}+|\phi_{1}|^{2}}) = N_{c}N_{f}\frac{p^{3}_{F}}{3\pi^{2}}, 
\end{eqnarray}
where, $\theta(x)$ is the step function defined as follows: $\theta(x)=1$ for $x>0$ and $\theta(x)=0$ for $x<0$.
$p_{F}$ is the Fermi momentum.
$\mu$ coincides with the Fermi energy $\sqrt{p^{2}_{F}+|\phi_{1}|^{2}}$ at $T=0$,
and it is determined by the baryon density of fermion. 
Later, we solve Eq. (44) numerically.
We completely neglect the temperature dependence of $\mu$, 
and use it as an external parameter for our numerical calculations.

\section{Quasiparticle Excitation Energy Spectra}

We will solve the gap equations (43) and (44) to study DCSB and CSC without BEC, in Sec. IV and Sec. V.
Before that, we examine the quasiparticle excitation energy spectra (30)-(37).
Because we consider the case $\mu\ne 0$,
we have to examine the possibility of BEC ( the condition for BEC ) of the scalar sector in our model. 
Depending on numerical values of the model parameters and gap functions,
these excitation branches $E^{B}_{+}$ and ${\cal E}^{B}_{+}$ can have zero-points. 
In such a situation, the logarithmic functions $\ln (1-e^{-\beta E^{B}_{+}})$ 
and/or $\ln (1-e^{-\beta {\cal E}^{B}_{+}})$ in $V_{eff}$ diverge, and the BEC takes place.
In this paper, we will not study the physical property of the BEC phase in our model,
and only examine the phase boundary in the model-parameter space.
This can be done by the examination of the quasiparticle energy spectra,
with neglecting the temperature dependence of $\mu$.

\vspace{3mm}

First, we examine the quasiparticle energy spectra (30)-(37) in several limiting cases.
The zero-density case at $\mu=0$, the energy spectra becomes
\begin{eqnarray}
E^{B}_{+}(\bmp) &=& E^{B}_{-}(\bmp) = \sqrt{\bmp^{2}+|\phi_{1}|^{2}+4|\phi_{2}|^{2}+\Delta^{2}},   \\
E^{F}_{+}(\bmp) &=& E^{F}_{-}(\bmp) = \sqrt{\bmp^{2}+|\phi_{1}|^{2}+4|\phi_{2}|^{2}},   \\
{\cal E}^{B}_{+}(\bmp) &=& {\cal E}^{B}_{-}(\bmp) = \sqrt{\bmp^{2}+|\phi_{1}|^{2}+\Delta^{2}},  \\
{\cal E}^{F}_{+}(\bmp) &=& {\cal E}^{F}_{-}(\bmp) = \sqrt{\bmp^{2}+|\phi_{1}|^{2}}.
\end{eqnarray}
On the other hand, when $|\phi_{2}|=0$, one finds the spectra of the case of the DCSB at finite density:
\begin{eqnarray}
E^{B}_{\pm}(\bmp) &=& {\cal E}^{B}_{\pm}(\bmp) = \sqrt{\bmp^{2}+|\phi_{1}|^{2}+\Delta^{2}}\mp\mu, \\
E^{F}_{\pm}(\bmp) &=& {\cal E}^{F}_{\pm}(\bmp) = \sqrt{\bmp^{2}+|\phi_{1}|^{2}}\mp\mu.
\end{eqnarray}
In the case only a CSC gap is generated, $|\phi_{1}|=0$, the spectra become:
\begin{eqnarray}
E^{B}_{\pm}(\bmp) &=& \sqrt{\bmp^{2}+4|\phi_{2}|^{2}+\Delta^{2}}\mp\mu, \\
E^{F}_{\pm}(\bmp) &=& \sqrt{(|\bmp|\mp\mu)^{2}+4|\phi_{2}|^{2}}, \\
{\cal E}^{B}_{\pm}(\bmp) &=& \sqrt{\bmp^{2}+\Delta^{2}}\mp\mu,  \\
{\cal E}^{F}_{\pm}(\bmp) &=& |\bmp|\mp\mu.
\end{eqnarray}
In Eqs. (51), (53) and (55), $E^{B}_{+}(\bmp)$ and ${\cal E}^{B}_{+}(\bmp)$ can be negative at $\mu\ne 0$.
Because of the positiveness of the Bose distribution functions 
$1/(e^{\beta E^{B}_{+}}-1)$ and $1/(e^{\beta {\cal E}^{B}_{+}}-1)$,
and $\mu$ corresponds to the Fermi energy of the system at zero temperature, 
\begin{eqnarray}
|\phi_{1}| \le \mu \le \sqrt{|\phi_{1}|^{2}+\Delta^{2}}
\end{eqnarray}
has to be satisfied in the case of Eqs. (51) and (52). 
By using $\mu=\sqrt{p^{2}_{F}+|\phi_{1}|^{2}}$, (57) will be rewritten as
the condition of the external parameters $p_{F}$ and $\Delta$:
\begin{eqnarray}
0 \le p_{F} \le \Delta.
\end{eqnarray} 
Thus, $\Delta$ is the upperbound for $p_{F}$ in the SUSY theory.
Similarly, for Eqs. (53)-(56),
\begin{eqnarray}
0 \le \mu=p_{F} \le \Delta
\end{eqnarray}
has to be satisfied.

\vspace{3mm}

In Fig. 1, we show a typical case of the excitation energy spectra of boson $E^{B}_{\pm}(\bmp)$ and 
fermion $E^{F}_{\pm}(\bmp)$ quasiparticles under the CSC state with a non-vanishing chiral mass $|\phi_{1}|$.
$E^{F}_{+}(\bmp)$ has a minimum at $p_{F}=\sqrt{\mu^{2}-|\phi_{1}|^{2}}$ 
where the excitation energy gap $4|\phi_{2}|$ locates.
The appearances of the branches $E^{B}_{\pm}(\bmp)$ and ${\cal E}^{B}_{\pm}(\bmp)$ are the new phenomenon
of our SUSY theory compared with the non-SUSY relativistic BCS (color)superconductivity~[6,18,19].
All of the spectra become parallel with the light cone at $|\bmp|\to\infty$.
In this figure, the Bose branches $E^{B}_{+}$ and ${\cal E}^{B}_{+}$ have energy gaps at $|\bmp|=0$:
\begin{eqnarray}
E^{B}_{+}(\bmp=0) &=& \sqrt{|\phi_{1}|^{2}+4|\phi_{2}|^{2}+\Delta^{2}}-\mu,  \\
{\cal E}^{B}_{+}(\bmp=0) &=& \sqrt{|\phi_{1}|^{2}+\Delta^{2}}-\mu.
\end{eqnarray}
Note that both $E^{B}_{+}(\bmp)$ and ${\cal E}^{B}_{+}(\bmp)$ depend on $\bmp^{2}$ as monotonically incleasing functions,
and
\begin{eqnarray}
E^{B}_{+}(\bmp) \ge {\cal E}^{B}_{+}(\bmp) \ge 0
\end{eqnarray}
is always satisfied ( ${\cal E}^{B}_{+}(\bmp) < 0$ is unphysical ). 
Thus, the BEC never take place in the branch $E^{B}_{+}$ at $|\phi_{2}|\ne 0$,
and the condition of BEC will be defined by ${\cal E}^{B}_{+}(\bmp=0)=0$.
From this equation, we find 
\begin{eqnarray}
\Delta=\sqrt{\mu^{2}-|\phi_{1}|^{2}}
\end{eqnarray}
as the condition of the BEC. 
Under this condition, ${\cal E}^{B}_{+}(\bmp)$ ( $=E^{B}_{+}(\bmp)$ when $|\phi_{2}|=0$ ) behaves at $|\bmp|\to 0$ as 
\begin{eqnarray}
{\cal E}^{B}_{+}(\bmp) &\approx& \frac{\bmp^{2}}{2\sqrt{(\Delta^{BEC}_{cr})^{2}+|\phi_{1}|^{2}}}.
\end{eqnarray}
Namely, there is no energy gap of the branch ${\cal E}^{B}_{+}$ under the condition (63). 
We conclude that, the BEC can take place at $\Delta=\sqrt{\mu^{2}-|\phi_{1}|^{2}}$,
and the BEC can coexist with the DCSB and the CSC. 
Always $0\le p_{F}\le \Delta$ has to be satisfied.
We should mention that, these results on the relations between the BEC, DCSB and CSC 
depend on the breaking scheme of the gauge and flavor symmetries. 
The results given above are for the breaking scheme of the $SU(3_{c})\times U(2_{f})_{L}\times U(2_{f})_{R}$ case
discussed in the previous section.

\section{Dynamical Chiral Symmetry Breaking}

In this section, we study DCSB without CSC ( $|\phi_{1}|\ge 0$, $|\phi_{2}|=0$ ) and BEC.

\subsection{${\cal N}=1$, four-dimensional case}

First, we consider the ${\cal N}=1$ case.
At $T=0$ and $|\phi_{2}|=0$, the gap equation (43) becomes
\begin{eqnarray}
1 &=& \frac{2G_{1}N_{c}N_{f}}{\pi^{2}}\int^{\Lambda}_{0}p^{2}dp\Bigl( \frac{\theta(p-p_{F})}{\sqrt{p^{2}+|\phi_{1}|^{2}}} -\frac{1}{\sqrt{p^{2}+|\phi_{1}|^{2}+\Delta^{2}}} \Bigr)  \nonumber \\
&=& \frac{G_{1}N_{c}N_{f}}{\pi^{2}}\Bigl( \Lambda\sqrt{\Lambda^{2}+|\phi_{1}|^{2}}-|\phi_{1}|^{2}\ln\frac{\Lambda+\sqrt{\Lambda^{2}+|\phi_{1}|^{2}}}{|\phi_{1}|}   \nonumber \\
& & \qquad -\Lambda\sqrt{\Lambda^{2}+|\phi_{1}|^{2}+\Delta^{2}} + (|\phi_{1}|^{2}+\Delta^{2})\ln\frac{\Lambda+\sqrt{\Lambda^{2}+|\phi_{1}|^{2}+\Delta^{2}}}{\sqrt{|\phi_{1}|^{2}+\Delta^{2}}} \nonumber \\
& & \qquad -p_{F}\sqrt{p^{2}_{F}+|\phi_{1}|^{2}}+|\phi_{1}|^{2}\ln\frac{p_{F}+\sqrt{p^{2}_{F}+|\phi_{1}|^{2}}}{|\phi_{1}|}   \Bigr).
\end{eqnarray} 
This equation is the same with the $U(1)$ case except the difference of the coefficient of $G_{1}$~[16].
At $\Delta\to 0$, the right-hand side becomes negative, thus there is no nontrivial solution.
With taking the limit $\Delta\to \infty$, 
Eq. (65) gives the gap equation of the DCSB of the non-SUSY NJL model at finite density~[7], 
while at $p_{F}=0$ it will give the gap equation of the zero-density case of the SNJL model~[4,15].
To see the relation of several possible phases in the space of parameters
$G_{1}N_{c}N_{f}\Lambda^{2}$, $\Delta/\Lambda$ and $p_{F}/\Lambda$, we examine of the critical coupling for the DCSB.
The determination equation for the critical coupling $(G_{1})_{cr}$ is obtained as follows:
\begin{eqnarray}
(G_{1})_{cr}N_{c}N_{f}\Lambda^{2} &=& \frac{\pi^{2}}{1-\sqrt{1+\frac{\Delta^{2}}{\Lambda^{2}}}+\frac{\Delta^{2}}{\Lambda^{2}}\ln\frac{1+\sqrt{1+\Delta^{2}/\Lambda^{2}}}{\Delta/\Lambda}-\frac{p^{2}_{F}}{\Lambda^{2}}}.
\end{eqnarray}
Figure 2 shows $(G_{1})_{cr}N_{c}N_{f}\Lambda^{2}$ as a function of $\Delta/\Lambda$.
$(G_{1})_{cr}N_{c}N_{f}\Lambda^{2}$ of the non-SUSY case is obtained by 
$\lim_{\Delta/\Lambda\to\infty}(G_{1})_{cr}N_{c}N_{f}\Lambda^{2}=\pi^{2}/(1-p^{2}_{F}/\Lambda^{2})$.
The denominator of Eq. (66) includes the finite-density effect on $(G_{1})_{cr}$.
$(G_{1})_{cr}$ of $p_{F}\ne 0$ is larger than that of $p_{F}=0$.
The divergence of $(G_{1})_{cr}N_{c}N_{f}\Lambda^{2}$ at 
a non-zero value of $\Delta/\Lambda$ ( depends on a numerical value of $p_{F}$ )
indicates the existence of the critical soft mass $\Delta^{DCSB}_{cr}$ for the DCSB of this model~[14]. 
From (58), $\mu=p_{F} < \Delta$ has to be satisfied in Fig. 2:
When $p_{F}=\Delta$, the BEC takes place at the mode of $\bmp=0$.

\subsection{${\cal N}=2$, three-dimensional case}

In ${\cal N}=1$ ${\rm SQED}_{4}$ ( four-dimensional supersymmetric quantum electrodynamics ), 
it was concluded that the dynamical mass generation never occurs by solving the Schwinger-Dyson equation, 
while it can take place with introducing a SUSY-breaking soft mass~[25].
The similar situation happens in ${\cal N}=2$ ${\rm SQED}_{3}$ ( three-dimensional SQED ) 
obtained by a dimensional reduction of ${\rm SQED}_{4}$. 
In ${\rm QED}_{3}$, there are two-component fermion model and four-component fermion model~[26].
The ${\cal N}=2$ ${\rm SQED}_{3}$ obtained by a dimensional reduction of ${\cal N}=1$ ${\rm SQED}_{4}$ 
is the SUSY counterpart of the four-component-spinor version of ${\rm QED}_{3}$~[27,28,29].
In this paper, we follow the method of Ref.~[28], and assume all of the quantum fields do not depend
on one space-direction.
The dimensional reduction is done by the following substitution in our gap equations (43) and (44):
\begin{eqnarray}
\int \frac{dp_{3}}{2\pi} \to \frac{1}{r_{c}}, 
\end{eqnarray}
or, equivalently $\frac{1}{\pi}\int p^{2}dp \to \frac{1}{r_{c}}\int pdp$.
Here, we compactify the third direction ( $r_{c}$; compactification scale ).
Now, the gap equation (43) at $T=0$ and $|\phi_{2}|=0$ becomes
\begin{eqnarray}
1 &=& \frac{2G_{1}N_{c}N_{f}}{\pi r_{c}}\int^{\Lambda}_{0}pdp\Bigl( \frac{\theta(p-p_{F})}{\sqrt{p^{2}+|\phi_{1}|^{2}}} -\frac{1}{\sqrt{p^{2}+|\phi_{1}|^{2}+\Delta^{2}}} \Bigr)  \nonumber \\
&=& \frac{2G_{1}N_{c}N_{f}}{\pi r_{c}}\Bigl( \sqrt{\Lambda^{2}+|\phi_{1}|^{2}} - \sqrt{p^{2}_{F}+|\phi_{1}|^{2}} - \sqrt{\Lambda^{2}+|\phi_{1}|^{2}+ \Delta^{2}} + \sqrt{|\phi_{1}|^{2}+\Delta^{2}} \Bigr).
\end{eqnarray}
Here, $\Lambda$ is a two-dimensional momentum cutoff.
At $\Delta\to 0$, there is no nontrivial solution similar to the case of the ${\cal N}=1$ four-dimensional theory.
The critical coupling will be determined by
\begin{eqnarray}
\frac{(G_{1})_{cr}N_{c}N_{f}\Lambda}{r_{c}} &=& \frac{\pi/2}{1-\sqrt{1+\frac{\Delta^{2}}{\Lambda^{2}}}+\frac{\Delta}{\Lambda}-\frac{p_{F}}{\Lambda} }.
\end{eqnarray}
Again, we find that the finite-density effect suppresses the DCSB similar to the ${\cal N}=1$ four-dimensional case.
The non-SUSY case is obtained by 
$\lim_{\Delta/\Lambda\to\infty}(G_{1})_{cr}N_{c}N_{f}\Lambda/r_{c}=(\pi/2)/(1-p_{F}/\Lambda)$.
$(G_{1})_{cr}N_{c}N_{f}\Lambda/r_{c}$ diverges at $\Delta\to 0$, $p_{F}\to 0$.
Figure 3 shows $(G_{1})_{cr}N_{c}N_{f}\Lambda/r_{c}$ as a function of $\Delta/\Lambda$.
In this figure, $p_{F}<\Delta$ has to be satisfied.
The dependence of $(G_{1})_{cr}N_{c}N_{f}\Lambda/r_{c}$ on $\Delta/\Lambda$ 
shows qualitatively the same behavior with the ${\cal N}=1$ four-dimensional case.

\section{Color Superconductivity at $N_{c}=3$, $N_{f}=2$}

In this section, we restrict ourselves to study a CSC state of the $SU(3_{c})\times U(2_{f})_{L}\times U(2_{f})_{R}$ model 
without DCSB ( $|\phi_{1}|=0$, $|\phi_{2}|\ge 0$ ) and BEC.
Usually, the superconducting energy gaps are much smaller than $\mu$ in various superconductors. 
Here, we assume it is also the case in our numerical calculation for solving Eq. (44),
and choose the model parameters $G_{2}$, $\Lambda$, $\mu$, and $\Delta$ to satisfy $2|\phi_{2}|/\mu \ll 1$.
The BCS gap function $2|\phi_{2}|$ is quite sensitive to these model parameters, 
we have to carefully choose the numerical values of them ( a kind of fine-tuning ) 
to get a physically reasonable solution:
For example, a too large $G_{2}$ easily gives a solution larger than $\mu$,
and it should be regarded as an unphysical one.
We also have to take into account $p_{F}<\Delta$ for choosing a numerical value of the soft mass
to avoid the condition for the realization of the BEC, because our theory cannot handle the BEC state
as a stable vacuum.

\subsection{${\cal N}=1$, four-dimensional case}

In the ${\cal N}=1$ four-dimensional case, our gap equation (44) is the same with the $U(1)$-gauge case~[16], 
except the coefficient of the coupling $G_{2}$. 
In Eq. (44), the coefficient $2N_{f}$ ( coming from the breaking scheme which depends on $N_{c}$ and $N_{f}$ ) 
is absorbed in $G_{2}$, and the numerical results are qualitatively equivalent with that of $U(1)$ case:
The numbers of $N_{c}$ and $N_{f}$ have no specific meaning in our gap equation of CSC.

\vspace{3mm}

Figure 4(a) shows the gap function $2|\phi_{2}|$ of the CSC state at $T=0$,
as a function of the coupling constant $G_{2}$.  
Equation (44) was solved under the five examples: $\Delta/\Lambda=0.45$, $0.8$, $2$, $10$, and the non-SUSY case. 
The non-SUSY case, obtained by dropping the contribution coming from the bosonic part in Eq.(44), 
corresponds to the ordinary $SU(3_{c})\times U(2_{f})_{L}\times U(2_{f})_{R}$ CSC.
The gap function $2|\phi_{2}|$ of all examples show qualitatively similar dependence on $G_{2}$, 
whether a numerical value of $\Delta$ is larger or smaller than $\Lambda$, 
and they reflect the nonperturbative dependence on $G_{2}$.
In all of the cases, the effect of SUSY suppresses the magnitude of CSC gap functions.
When the soft mass becomes large, the dependence on $G_{2}$ becomes strong.
The result in Fig. 4(a) also indicates the absence of the critical coupling 
of the CSC similar to the ordinary BCS theory~[1]:
Any attractive interaction gives a superconducting instability also in the SUSY CSC.

\vspace{3mm}

In Fig. 4(b), we show the gap function $2|\phi_{2}|$ at $T=0$ as a function of $\Lambda$
under the energy unit $\mu=1$. When $\Delta$ takes a value closed with $\mu$, 
the divergence of the gap function becomes moderate compared with other examples because of the effect of SUSY:
$2|\phi_{2}|$ diverges almost linearly.
When $\Delta$ is large enough, $2|\phi_{2}|$ will diverge almost quadratically under $\Lambda\to \infty$.

\vspace{3mm}

Figure 4(c) gives $2|\phi_{2}|$ as a function of temperature $T$.
The gap function continuously vanishes at $T\to T_{c}$ ( $T_{c}$; the critical temperature ),
clearly shows the character of second-order phase transition of the superconductivity.
In all of the examples shown in this figure, the BCS universal constant
$2|\phi_{2}(T=0)|/T_{c}=1.76$ is satisfied~[1]. Both $2|\phi_{2}(T=0)|$ and $T_{c}$ depend on 
$G_{2}$, $\Lambda$, $\mu$ and $\Delta$, though the ratio of them is independent on these parameters: 
The effect of these model parameters are canceled with each other by taking the ratio. 
In the non-relativistic~[1], the relativistic~[19], 
and the $U(1)$ SUSY~[16] BCS theories, any spin-singlet Lorentz-scalar BCS gap function satisfies the universal constant.
We find it is also satisfied in our $SU(3_{c})$ SUSY CSC theory.

\vspace{3mm}

We discuss the thermodynamic property of the SUSY CSC.
The definitions of the entropy $S$ and the heat capacity $C$ are given by
\begin{eqnarray}
S \equiv -\frac{\partial V_{eff}}{\partial T}, \qquad 
C \equiv T\frac{\partial S}{\partial T}.
\end{eqnarray}
In our case, Eq. (42) of $V_{eff}$ will be used to obtain $S$ and $C$.
In our model, the quasiparticle excitation branches will be classified into two classes:
(A) $E^{B}_{\pm}$, $E^{F}_{\pm}$, and (B) ${\cal E}^{B}_{\pm}$, ${\cal E}^{F}_{\pm}$.
When the BEC does not take place, all of the branches $E^{B}_{\pm}$, $E^{F}_{\pm}$, ${\cal E}^{B}_{\pm}$
and ${\cal E}^{F}_{-}$ have energy gaps, while only ${\cal E}^{F}_{+}$ is gapless.
The thermodynamic characters of $S$ and $C$ are determined by simple sums of the contributions
of the ideal gas of quasiparticles of these branches.
To examine the contribution of the branches of (A) in thermodynamics, 
let us recall the energy spectra shown in Fig. 1.
Because the density of states diverges around the energy gap of $E^{F}_{+}$~[19], 
thermal excitations of $E^{F}_{+}$ dominate the contribution of the class (A) in the integrals of $S$ and $C$, 
and other branches of (A) give almost no contribution: 
Namely, the contributions of both the entropy $S$ and the heat capacity $C$ coming from (A) are determined 
by the thermal excitations of the quasiparticles of the branch $E^{F}_{+}$. 
Because the temperature dependence of $2|\phi_{2}|$ is the same with the well-known BCS result, 
the contributions of thermal excitations of the branches of (A) 
for $S$ and $C$ are suppressed by the factor $\exp[-2|\phi_{2}(T)|/T]$: 
They will show the same behavior with the usual BCS theory. 
On the other hand, the low-temperature thermodynamics in the branches of (B) 
is determined by the gapless excitation ${\cal E}^{F}_{+}$, and it is just the free Fermi gas in our model.
Similar to the case of CSC in ordinary QCD, the total thermodynamics of the system 
is given by the superposition of the contribution of the quasiparticles of (A) and (B).
As a consequence, we conclude that the thermodynamic character of 
the SUSY CSC without BEC is the same with the non-SUSY CSC.

\subsection{${\cal N}=2$, three-dimensional case}

We examine the CSC gap equation (44) with the dimensional reduction (67).
Figure 5(a) shows the CSC gap function $2|\phi_{2}|$ of the ${\cal N}=2$ three-dimensional case 
at $T=0$ as a function of $G_{2}/r_{c}$. The behavior is almost the same with ${\cal N}=1$ four-dimensional
case: No critical coupling, nonperturbative dependence on $G_{2}$, and suppression of $2|\phi_{2}|$ 
by a non-vanishing $\Delta$.

\vspace{3mm}

Figure 5(b) shows $2|\phi_{2}|$ at $T=0$ as a function of $\Lambda$, with the energy unit $\mu=1$.
In the case of $\Delta/\mu=1.125$, the divergence of $2|\phi_{2}|$ is quite slow ( almost logarithmically ),
while $2|\phi_{2}|$ linearly diverges in the $\Delta/\mu=5$ case. 
The non-SUSY case diverges almost quadratically at $\Lambda\to\infty$.
These differences of the order ( quadratic, linear, logarithmic ) of the divergence of $2|\phi_{2}|$ 
in several examples are determined by both the dimensions of space and a numerical value of $\Delta/\mu$.

\vspace{3mm}

In Fig. 5(c), we show $2|\phi_{2}|$ as a function of $T$.
The result is qualitatively similar with Fig. 4(c):
$2|\phi_{2}|$ continuously vanishes at $T\to T_{c}$, and the ratio $2|\phi_{2}(T=0)|/T_{c}=1.76$ is always satisfied. 
In fact, this ratio is always satisfied in the three-dimensional non-SUSY BCS theory~[30].
We find it is also the case in the ${\cal N}=2$ three-dimensional SUSY BCS CSC theory.

\section{Concluding Remarks}

In summary, we have discussed the BEC, DCSB and CSC 
in the $SU(N_{c})\times U(N_{f})_{L}\times U(N_{f})_{R}$-invariant generalized SNJL model
at finite temperature and density.
The condition for the realization of the BEC has been examined 
especially in the example of the $SU(3_{c})\times U(2_{f})_{L}\times U(2_{f})_{R}$ case.
The gap equations for the DCSB and the CSC have been derived, and have been solved
for both the ${\cal N}=1$ four-dimensional and the ${\cal N}=2$ three-dimensional cases.
We have found the finite-density effect in the critical coupling of the DCSB in both cases.
The effects of the bosonic part of the gap equation (44) in the SUSY BCS-type CSC theory have been reviewed in detail. 
In the CSC, the SUSY effects in the gap function $2|\phi_{2}|$ become significant 
when $\Delta$ takes a numerically closed value to $\mu$.
Similar to the $U(1)$ gauge case of Ref.~[16], the CSC shows the BCS character even when $\mu$ is close to $\Delta$:
There is no critical coupling in the SUSY CSC, and the BCS universal constant $2|\phi_{2}(T=0)|/T_{c}=1.76$
is always satisfied both in the ${\cal N}=1$ four-dimensional and the ${\cal N}=2$ three-dimensional cases.
We should emphasized that, these results are nontrivial and indicate the universality of the BCS-NJL mechanism,
has nothing to do with the numerical value of $\mu$ ( from eV to TeV ), 
dimension of spacetime ( 3D or 4D ), SUSY or non-SUSY.

\vspace{3mm}

Finally, we would like to make some comments on several issues.
It is interesting for us to examine the collective modes in the DCSB and the CSC of our SUSY case at finite density. 
The collective modes or the Nambu-Goldstone modes have been studied in the non-SUSY CSC theory,
and an anomalous phenomenon of the Nambu-Goldstone theorem was found~[31].
An interaction between scalars might alter the excitation energy spectra $E^{B}_{\pm}$ and ${\cal E}^{B}_{\pm}$ of the boson sector,
as discussed in the Bogoliubov theory of superfluidity~[32].
For the examination of this problem, we should take into account the quantum fluctuations around a stationary point of $V_{eff}$
in our model.

\vspace{3mm}

In this paper, we examined the simplest example of the breaking scheme of $SU(3_{c})$ CSC.
However, from a phenomenological point of view, the results cannot be regarded seriously. 
Our choice of $SU(3_{c})$ gauge symmetry does not have a strong meaning, 
and should be considered as for a convenience of both the examination of the quasiparticle excitation 
energy spectra and the numerical calculations of the CSC gap equation (44). 
A more detailed examination of the CSC states under several numbers of $N_{c}$ and $N_{f}$ is an important issue~[24].
Moreover, our SNJL-type model does not correspond to a naive supersymmetric extension of the
QCD-motivated NJL model usually used as a low-energy effective theory of QCD. 
The most important aim of this work is the presentation of a method for an examination
of the BEC, DCSB and CSC in a SUSY condensed matter of a non-Abelian gauge model at finite density and temperature.
In the non-SUSY CSC, the $SU(3_{c})$ gauge interaction is the origin of the attractive interaction,
and it causes the Cooper instability. 
Therefore, if one uses an NJL-like model to describe the CSC, 
the model parameters of it should be chosen from the consideration
of the running coupling of QCD gauge interaction and hadron phenomenology. 
A coupling constant for CSC should take a numerical value of ${\cal O}({\rm GeV}^{-2})$, 
while a cutoff will become ${\cal O}({\rm GeV})$~[6,7],
and they are determined by a numerical value of the chiral condensate given by 
$\langle\bar{u}u\rangle$, $\langle \bar{d}d\rangle$ and the decay constant of pion.
At the order of this energy scale of the cutoff, there is no SUSY effect in the CSC 
in the sense of the context of this paper.
From the energy scale of $\Delta$ ( $>$TeV ) relevant for particle phenomenology, 
our theory is relatively closer to the electroweak theory, 
especially the top condensation model~[3,4] or the technicolor theory~[5] than QCD.
As a consequence, it is interesting for us to apply of our theory to several technicolor gauge groups.
In such a case, the situation $\mu \ge \Lambda_{TC}$ should be considered in our method.
The extensions to several gauge groups of SUSY grand unified theories ( SGUTs ) are also possible~[33,34].

\begin{figure}

\caption{The excitation energy spectra of quasiparticles of the branches $E^{B}_{+}(\bmp)$, 
$E^{B}_{-}(\bmp)$, $E^{F}_{+}(\bmp)$ and $E^{F}_{+}(\bmp)$ under the CSC state with non-vanishing chiral mass. 
We choose $\mu=2$, $|\phi_{2}|=0.02$, $\Delta=2$ and $|\phi_{1}|=1$.}

\caption{The critical coupling $(G_{1})_{cr}N_{c}N_{f}\Lambda^{2}$ of the DCSB 
in the ${\cal N}=1$ four-dimensional case, shown as a function of $\Delta/\Lambda$.}

\caption{The critical coupling $(G_{1})_{cr}N_{c}N_{f}\Lambda/r_{c}$ of the DCSB in 
the ${\cal N}=2$ three-dimensional case, shown as a function of $\Delta/\Lambda$.}

\caption{The CSC gap function $2|\phi_{2}|$ of the ${\cal N}=1$ four-dimensional 
$SU(3_{c})\times U(2_{f})_{L}\times U(2_{f})_{R}$ model. 
(a) At $T=0$, shown as a function of $G_{2}$. We set $\mu=0.4$ and $\Lambda=1$.
(b) At $T=0$, given as a function of $\Lambda$. We set $\mu=0.4$ and $G_{2}=1$.
(c) Given as a function of $T$. The model parameters are set as $\mu=0.4$, $\Lambda=1$, $G_{2}=1$.}

\caption{The CSC gap function $2|\phi_{2}|$ of the ${\cal N}=2$ three-dimensional 
$SU(3_{c})\times U(2_{f})_{L}\times U(2_{f})_{R}$ model. 
(a) At $T=0$, shown as a function of $G_{2}/r_{c}$. We set $\mu=0.4$ and $\Lambda=1$.
(b) At $T=0$, given as a function of $\Lambda$. We set $\mu=0.4$ and $G_{2}/r_{c}=0.2$.
(c) Given as a function of $T$. The model parameters are set as $\mu=0.4$, $\Lambda=1$, $G_{2}/r_{c}=0.18$.}

\end{figure}

\end{document}